\documentclass[10pt, twocolumn, nofootinbib,superscriptaddress]{revtex4-1}
\usepackage{amsmath,amssymb,amsfonts}
\usepackage{algorithmic}
\usepackage{graphicx}
\usepackage{textcomp}
\usepackage{xcolor}
\usepackage{ragged2e}
\usepackage{booktabs, makecell, tabularx}

\usepackage{gensymb}

\makeatletter
    \renewcommand\@make@capt@title[2]{%
     \@ifx@empty\float@link{\@firstofone}{\expandafter\href\expandafter{\float@link}}%
      {\textbf{#1}}\@caption@fignum@sep#2\quad}%
\makeatother
\makeatletter 
\renewcommand{\fnum@figure}{\textbf{Fig.~\thefigure}}
\makeatother
\usepackage{svg}

\usepackage{xcolor}

\def\BibTeX{{\rm B\kern-.05em{\sc i\kern-.025em b}\kern-.08em
    T\kern-.1667em\lower.7ex\hbox{E}\kern-.125emX}}
\begin{document}
\title{Surface acoustic waves Brillouin photonics on a silicon nitride chip}

\author{Yvan~Klaver}
\author{Randy~te~Morsche}
\author{Roel~A.~Botter}
\affiliation{Nonlinear Nanophotonics, MESA+ Institute of Nanotechnology, University of Twente, Enschede, the Netherlands}
\author{Batoul~Hashemi}
\author{Bruno~L.~Segat~Frare}
\affiliation{Department of Engineering Physics, McMaster University, 1280 Main Street West, Hamilton, ON L8S 4L7, Canada}
\author{Akhileshwar~Mishra}
\author{Kaixuan~Ye}
\affiliation{Nonlinear Nanophotonics, MESA+ Institute of Nanotechnology, University of Twente, Enschede, the Netherlands}
\author{Hamidu~Mbonde}
\author{Pooya~Torab~Ahmadi}
\author{Niloofar~Majidian~Taleghani}
\author{Evan~Jonker}
\affiliation{Department of Engineering Physics, McMaster University, 1280 Main Street West, Hamilton, ON L8S 4L7, Canada}
\author{Redlef~B.G.~Braamhaar}
\affiliation{Nonlinear Nanophotonics, MESA+ Institute of Nanotechnology, University of Twente, Enschede, the Netherlands}
\author{Ponnambalam~Ravi~Selvaganapathy}
\affiliation{Department of Mechanical Engineering and the School of Biomedical Engineering, McMaster University, Hamilton, ON L8S 4L7, Canada}
\author{Peter~Mascher}
\affiliation{Department of Engineering Physics, McMaster University, 1280 Main Street West, Hamilton, ON L8S 4L7, Canada}
\author{Peter~J.M.~van~der~Slot}
\affiliation{Nonlinear Nanophotonics, MESA+ Institute of Nanotechnology, University of Twente, Enschede, the Netherlands}
\author{Jonathan~D.B.~Bradley}
\affiliation{Department of Engineering Physics, McMaster University, 1280 Main Street West, Hamilton, ON L8S 4L7, Canada}
\author{David~Marpaung}
\email{Corresponding author: david.marpaung@utwente.nl}
\affiliation{Nonlinear Nanophotonics, MESA+ Institute of Nanotechnology, University of Twente, Enschede, the Netherlands}
%\affiliation{Department of Engineering Physics, McMaster University, 1280 Main Street West, Hamilton, ON L8S 4L7, Canada}

% Comment of Jon on the summary is a suggestion to add a motivation for SBS incorperation before moving to the platform problem statement.
\begin{abstract}
Seamlessly integrating stimulated Brillouin scattering (SBS) in a low-loss and mature photonic integration platform remains a complicated task. Virtually all current approaches fall short in simultaneously achieving strong SBS, low losses, and technological scalability. In this work we incorporate stong SBS into a standard silicon nitride platform by a simple deposition of a tellurium oxide layer, a commonly used material for acousto-optic modulators. In these heterogeneously integrated waveguides, we harness novel SBS interactions actuated by surface acoustic waves (SAWs) leading to more than two orders of magnitude gain enhancement. Three novel applications are demonstrated in this platform: (i) a silicon nitride Brillouin amplifier with 5 dB net optical gain, (ii) a compact intermodal stimulated Brillouin laser (SBL) capable of high purity radio frequency (RF) signal generation with 7~Hz intrinsic linewidth,  and (iii) a widely tunable microwave photonic notch filter with ultra-narrow linewidth of 2.2~MHz enabled by Brillouin induced opacity. These advancements can unlock an array of new RF and optical technologies to be directly integrated in silicon nitride.

\end{abstract}

\maketitle

\section*{Introduction}

\begin{figure*} [ht!]
\centering
\includegraphics[width=\textwidth]{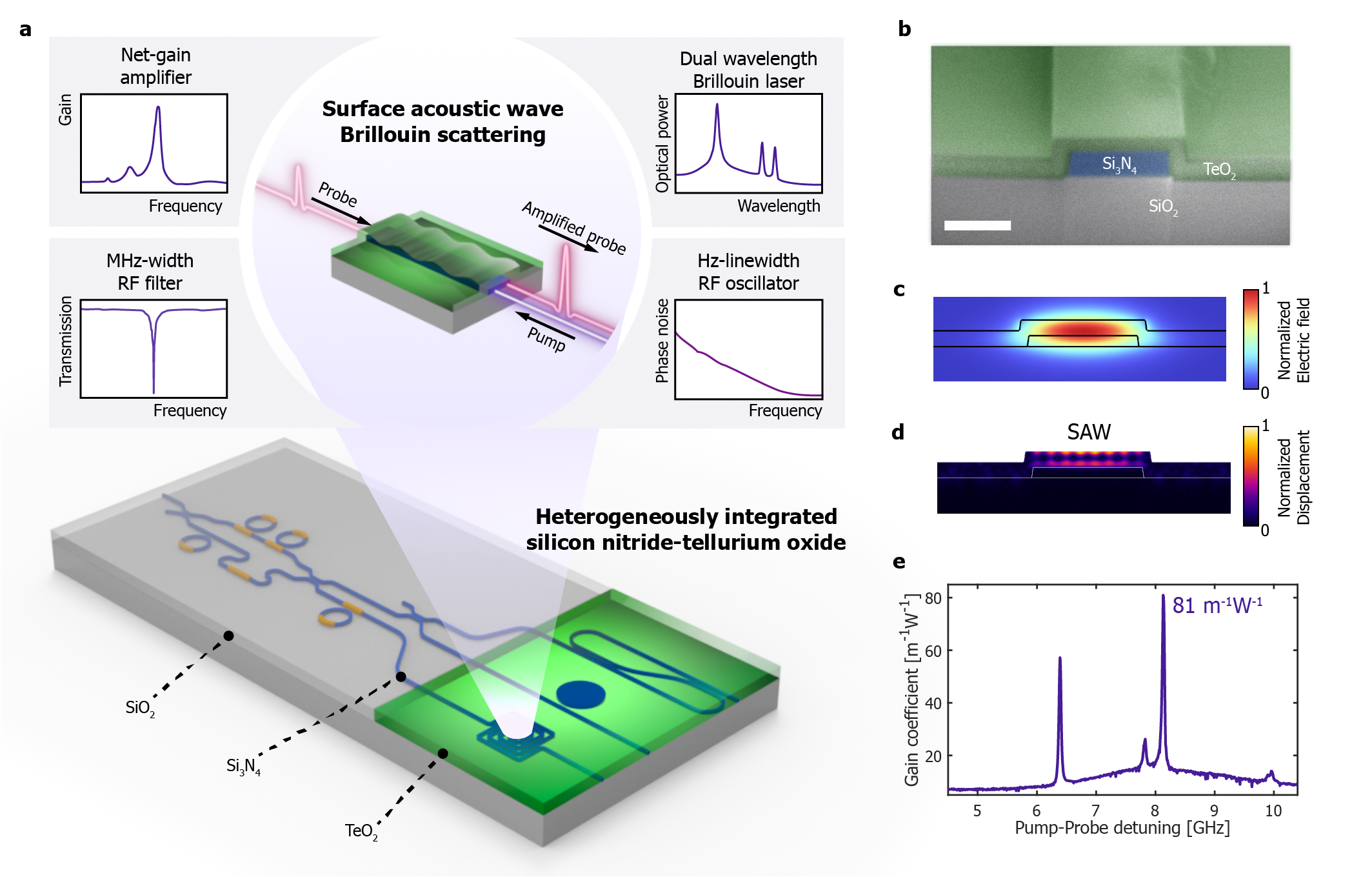} 
\caption{\textbf{Surface acoustic wave (SAW) based Stimulated Brillouin scattering in tellurite covered silicon nitride waveguides.} \textbf{a} Top centre: concept of Backwards stimulated Brillouin scattering mediated via a SAW mode supported in tellurite covered silicon nitride. Surrounding blocks: highlights of demonstrated capabilities in the SAW based silicon nitride platform.  Bottom: Concept art of hybrid integration of complex low loss silicon nitride circuitry with local activation of tellurite for Brillouin or rare-earth doping. \textbf{b} False colour SEM image of a tellurite covered silicon nitride waveguide cross-section. Scale bar: 1~µm \textbf{c} Finite element method (FEM) simulation result of the electric field distribution of the optical mode in the unclad waveguide. \textbf{d} FEM-simulated result of the displacement distribution of the acoustic response generated at the main Brillouin gain peak of e. \textbf{e} Measured Brillouin gain coefficient of the unclad waveguide shown in c,d.}\label{fig1}
\end{figure*}

Stimulated Brillouin scattering (SBS) \cite{Eggleton_NatPhot_2019,Merklein_APR_2022} is a coherent coupling between traveling photons and phonons that enables a wide range of technologically important functions including ultrahigh performance microwave photonic (MWP) filters \cite{Marpaung_Optica_2015}, narrow linewidth lasers \cite{Gundavarapu_NatPhot_2018}, RF oscillators \cite{Li_ncomms_2013,Heffernan_nphot_2024}, tunable delay lines \cite{Okawachi_PRL_2005,Chin_OE_2010}, and narrowband amplifiers \cite{Pelusi_OE_2017}. Unique characteristics of SBS, including optical gain, non-reciprocity \cite{Kim_nphys_2015,Sohn_NatPhot_2018,Kittlaus_NatPhot_2018}, and a very narrow linewidth resonance, are 
difficult to achieve in mainstream integrated photonics without extreme loss engineering. For this reason, integrating SBS in low-loss scalable platforms has thus far proven to be an insurmountable task.

A wide range of structures have been developed to achieve integration of SBS \cite{Eggleton_NatPhot_2019,Botter_SciAdv_2022,Lei_ncomms_2024}, but all come with significant drawbacks. Silicon SBS devices \cite{Kittlaus_NatPhot_2016, Otterstrom2018ALaser,Lei_ncomms_2024} lack in scalability due to the need of suspended structures to combat acoustic leakages. Approaches using chalcogenide soft glasses \cite{Pant_OE_2011,Marpaung_Optica_2015,Choudhary_JLT_2017} and their integration with silicon-on-insulator \cite{Morrison_Optica_2017,Matthew_ncomms_2023} promise strong gain but they suffer from the inherent material instability and toxicity of chalcogenide as well as the limiting nonlinear losses of silicon. Silicon-nitride-based platforms \cite{Gundavarapu_NatPhot_2018, Botter_SciAdv_2022, Ji_Science_2024, Ye_APL_2023, Ye_OMEX_2024, Gyger_PRL_2020,ZERBIB_result_2023}, on the other hand, promise ultra-low optical losses, but they currently suffer from inherent weak SBS gain, depriving a wide range of unique SBS functionalities from this important photonic platform. 

In this work, we explore the heterogeneous integration of silicon nitride circuits with tellurium oxide (TeO$_2$), also commonly known as tellurite, as a scalable and low-loss platform for strong Brillouin scattering that is free from all the drawbacks that have plagued previous demonstrations. Tellurite is a soft glass widely used commercially for acousto-optic modulators (AOMs) and has previously been investigated as an SBS material in bulk \cite{Yano_JApplPhys_1971} and fiber \cite{Abedin_OE_2006} forms. We employ a simple deposition step to add tellurite glass onto foundry-fabricated silicon nitride circuits and we unlock a novel SBS interaction mediated by surface acoustic waves (SAWs). 

Surface acoustic wave SBS has recently attracted significant attention due to its potential of achieving stronger Brillouin interactions with lower acoustic frequencies that exhibit lower losses. Initially studied in subwavelength fiber tapers and photonic crystal fibers \cite{Beugnot_ncomms_2014,Tchahame_OL_2016,Xu_optica_2023}, with potential applications in acoustic sensing, recently new works are reporting observation of SAW SBS in integrated platforms \cite{Neijts_Arxiv_2023,Ye_Arxiv_2023,rodrigues_arxiv_2023,Munk_ncomms_2019,Iyer_ncomms_2024}. Here we harness SAW in the tellurite layer to achieve two-orders of magnitude SBS enhancement in our silicon nitride circuit \cite{Botter_SciAdv_2022}.   

With our novel SAW SBS platform, we demonstrate an array of SBS functionalities previously inaccessible in silicon nitride (see Fig.\ref{fig1}a), including a Brillouin amplifier exhibiting a net gain that can be harnessed for RF filtering and true time delay. We then devoloped a compact 500~µm-diameter planar disk resonator to demonstrate intermodal Brillouin lasing with tunable single and dual-wavelength emission that operates in the phonon lasing regime. The prior leads to the generation of an RF signal with exceptional narrow linewidth of 7~Hz. We explore non-reciprocity in the micro-resonator through Brillouin-induced transparency and opacity, and in the process demonstrated an ultra-selective RF photonic tunable notch filter with a linewidth of 2.2~MHz. 

Our results promise significant improvements and, at the same time, expand new integrated photonic functions that can directly be embedded in a low-loss and scalable platform. These include low phase noise microwave synthesizers \cite{Li_ncomms_2013,Heffernan_nphot_2024}, Brillouin-Kerr microcombs \cite{Bai_PRL_2021}, ultra-precision photonic thermometry \cite{Loh_optica_2019,Lai_optica_2022}, and RF photonic signal processors \cite{Botter_SciAdv_2022}. Our approach also allows intersecting these Brillouin-based technologies with rare-earth-doped optical amplifiers \cite{Frankis_PhotRes_2020}, pump lasers \cite{Kiani_OE_2023}, as well as Kerr nonlinear signal processing \cite{Mbonde_FIO_2022} readily available in this versatile heterogeneous platform.

\section*{Results}\label{sec2}

\subsection*{Surface acoustic wave SBS}
Strong opto-acoustic overlap is of utmost importance for unlocking low-power chip-based SBS devices. To date, virtually all SBS demonstrations in silicon nitride suffer from poor confinement of bulk acoustic waves generated in the core region, significantly hindering the overlap. This leads to an impractically low SBS gain coefficient of 0.1-0.4~m$^{-1}$W$^{-1}$ \cite{Gundavarapu_NatPhot_2018,  Ji_Science_2024, Gyger_PRL_2020}, which is 3--4 orders of magnitude lower to that of silicon devices \cite{Kittlaus_NatPhot_2016,Otterstrom2018ALaser,Lei_ncomms_2024}. In contrast, our device exploits a SAW localized on the surface of the tellurite layer added to the silicon nitride waveguides (see Fig.\ref{fig1}b,c,d) allowing for enhanced overlap and at the same time accessing the high photo-elastic coefficient of tellurite ($p_{12} = 0.24$) \cite{Yano_JApplPhys_1971} which is 5 times stronger than that of silicon nitride \cite{Gyger_PRL_2020}. 

\begin{figure*} [ht!]
\centering
\includegraphics[width=\linewidth]{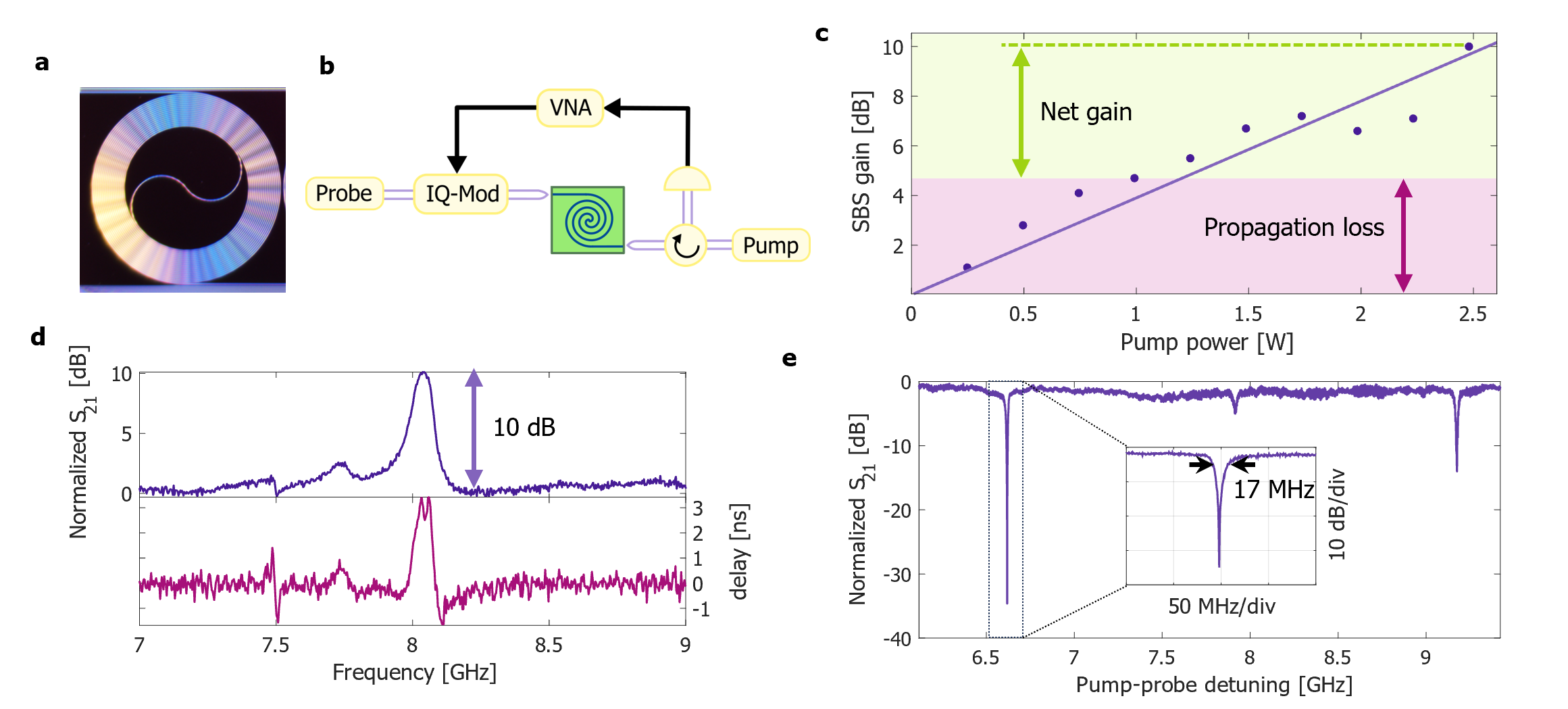}
\caption{\textbf{SBS applications in single pass waveguides.} \textbf{a} Photo of an 16~cm-long spiral waveguide used in the microwave photonic notch filter. \textbf{b} schematic of Vector Network Analyzer (VNA) based measurements used in the net-gain amplifier, delay and microwave photonic notch filter. IQ-mod: In-phase quadrature modulator. \textbf{c}  Measured on-off gain induced by SBS vs pump power, with a coloured region indicating the propagation loss in the same waveguide. The solid line represents a least squared fit to the data. \textbf{d} RF Transmission spectrum of the Brillouin based amplifier pumped at 2.5~W, with the corresponding group delay calculated from the phase response. \textbf{e} RF Transmission spectrum ($S_{21}$) of a microwave photonic notch filter response using the Brillouin gain.}\label{fig2}
\end{figure*}

We have performed extensive opto-acoustic optimization through geometry engineering using a genetic algorithm and cladding material engineering and focus on a particular cross-section without top-cladding that provides both enhanced Brillouin strength and fabrication robustness, see methods for dimensions. The structure supports two different types of guided acoustic modes, one higher-order SAW mode with maximum displacement at the surface (Fig.\ref{fig1}d) at an acoustic frequency of 8.13~GHz and one bulk acoustic mode with maximum displacement in the tellurite layer at 6.39~GHz. Both modes can mediate strong Brillouin interaction, in particular unlocking a novel SAW SBS interaction with a peak gain coefficient of 81~m$^{-1}$W$^{-1}$ (see Fig.\ref{fig1}e), which is a 200~times improvement over previous measured results in silicon nitride \cite{Gundavarapu_NatPhot_2018, Botter_SciAdv_2022, Ji_Science_2024, Ye_APL_2023, Ye_OMEX_2024, Gyger_PRL_2020,ZERBIB_result_2023}. Along with low propagation and nonlinear losses in our platform, this high gain directly allows us to demonstrate various SBS applications in waveguides and optical resonators.

\subsection*{Brillouin amplifier, delay line, and notch filter}
Integrating an optical amplifier directly into silicon nitride waveguides has recently recieved significant attention, the typical implementations are using rare-earth doping directly \cite{Liu_science_2022} or via heterogeneous integration of a host material \cite{Gaafar_ncomms_2024,Frankis_PhotRes_2020}. Here combining the strong Brillouin amplification and low loss we demonstrate (see methods and Fig.\ref{fig2}b) for the first time net-gain Brillouin amplification in silicon nitride, where the Brillouin gain overcomes the total propagation loss. An on-chip gain of 10~dB was observed leading to 5~dB net gain as demonstrated in Fig.\ref{fig2}c, using a 6.7-cm-long waveguide. Note that due to the negligible of two-photon absorption in both tellurite and silicon nitride, we observe no clamping of the SBS gain. We expect that with lower propagation loss, the on-off gain can be improved to over 40~dB with the same pump power, this brings our system close to the performance of amplifiers demonstrated in chalcogenide waveguides \cite{Choudhary_JLT_2017}.

\begin{figure*}[ht]
\centering
\includegraphics[width=\textwidth]{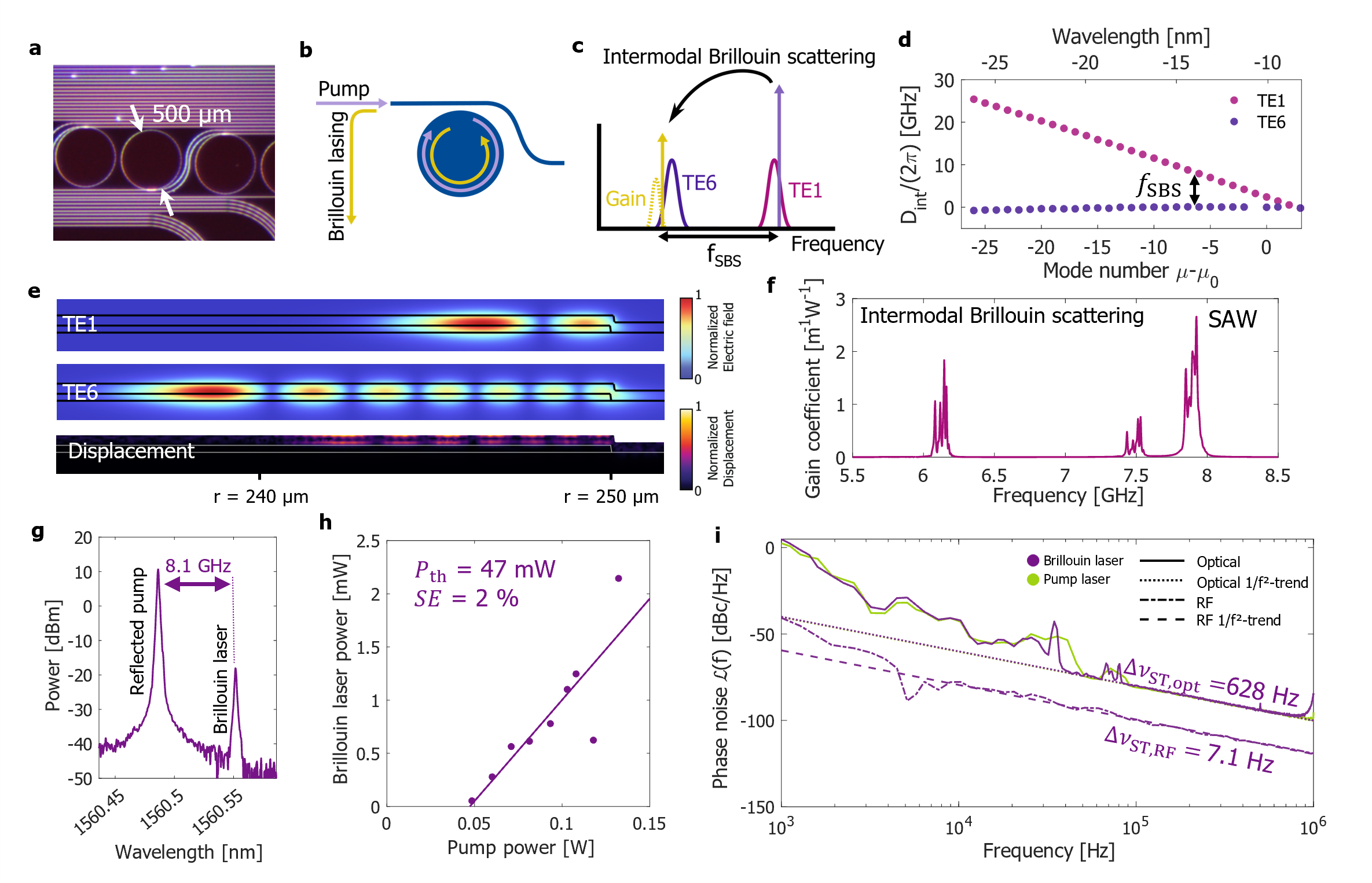}
\caption{\textbf{Microdisk resonator properties and Brillouin lasing.} \textbf{a} Photo of the microdisk. \textbf{b} Schematic of Brillouin lasing in a microdisk using backward Brillouin scattering. \textbf{c} Schematic of the resonance spectrum where TE1 is pumped and TE6 overlaps with the Brillouin gain to generate Stokes light. \textbf{d} Measured integrated cavity dispersion of the TE6 mode family, and frequency spacing to neighbouring mode family TE1. \textbf{e} Simulated optical mode profiles for the TE1 and TE6 mode in the disk, showing the normalized electric field and the acoustic response generated at the peak intermodal Brillouin gain displayed as the normalized displacement distribution. \textbf{f} Simulated intermodal Brillouin gain coefficient between mode TE1 and TE6. \textbf{g} Reflected optical spectrum recorded of the Brillouin laser. \textbf{h} On-chip generated laser power vs pump power with a least squared fit to extract the threshold power ($P_\mathrm{th}$) and slope efficiency (SE). \textbf{i} Phase noise $\mathcal{L}(f)$ measured for the Brillouin and pump laser and the beatnote between reflected pump and Brillouin laser. Intrinsic linewidths $\Delta\nu_\mathrm{ST}$ are extracted from the $f^{-2}$ fit.}\label{fig3}
\end{figure*}

Another important SBS application is microwave photonics \cite{Marpaung_NatPhot_2019} where the complex SBS resonances are used to process the phase and amplitude of microwave and optical signals \cite{Pant_lpr_2014}. Through the Kramers-Kronig relation, our Brillouin amplifier is capable to provide slow light group delay of 3.4~ns as depicted in Fig.\ref{fig2}d, comparable to what has been previously demonstrated in chalcogenide devices \cite{Aryanfar_OL_2017}. The narrow linewidth of the SBS response can also be harnessed for MWP filtering when the probe light is modulated with high frequency RF signals. By properly tailoring the phase and amplitude of the generated RF-modulation optical sidebands \cite{Marpaung_Optica_2015}, we transform the SBS gain resonance from a 16-cm-long waveguide with silica cladding (Fig.\ref{fig2}a, methods), into an ultra high rejection MWP notch filter with a 35~dB rejection and a narrow 17-MHz, -3dB linewidth (Fig.\ref{fig2}e). Due to the presence of multiple acoustic modes in our waveguide, the MWP filtering capability can also be extended to simultaneous filtering of multiple frequencies with a simple single-frequency pump wave. In contrast, a prior demonstration of multiple-band notch filtering required complex multi-pump arrangements \cite{Garrett_JLT_2022}.

\subsection*{Intermodal Brillouin lasing}
Inducing SBS in low-loss cavities can lead to a variety of advancements of important technology and physical observations including Brillouin lasers \cite{Gundavarapu_NatPhot_2018}, gyroscopes \cite{Gundavarapu_NatPhot_2018}, non-reciprocity \cite{Kim_nphys_2015}, optical frequency combs \cite{Bai_PRL_2021} and microwave synthesizers \cite{Li_ncomms_2013}. 
In general, the challenge to achieve Brilloun lasing stem from the difficulty to match the cavity modes with the Brillouin shift, that requires high precision due to the narrow linewidth of the SBS response, typically on the order of 10s of MHz. 
We overcome this challenge through intermodal Brillouin scattering between two transverse optical modes of a microdisk cavity. Intermodal SBS is typically used to enable optical gain in systems that only support forward Brillouin scattering \cite{Kittlaus_ncomms_2017,Otterstrom2018ALaser}. In our case, we took advantage of this effect to allow the SBS interaction to occur in an ultra small cavity, where the free spectral range (FSR) is much larger (91~GHz) than the Brillouin shift (8.1~GHz).

The principle of operation of our intermodal Brillouin laser is as follows, we map the cavity dispersion of multiple transverse optical modes belonging to a 500~µm-diameter TeO$_2$-Si$_3$N$_4$ microdisk (Fig.\ref{fig3}a), to find mode-pairs that can be coupled via Brillouin interactions (Fig.\ref{fig3}b,c). Figure~\ref{fig3}d depicts the measured integrated cavity dispersion of mode TE6 with neighbouring mode TE1, demonstrating that at around 1555~nm the mode spacing matches the SBS shift.
Through finite element method (FEM) simulation of the opto-acoustic interaction we predict the spatial distribution of the two optical modes along with the acoustic distribution (Fig.\ref{fig3}e) that can support SBS interaction via guided SAW. We compute the strength of this intermodal Brillouin scattering to be 2.7~m$^{-1}$W$^{-1}$ as depicted in Fig.\ref{fig3}f. 
While this is reduced from the strength of intramodal SBS scattering observed in a single pass waveguide, the high optical finesse (180) of our cavity still allows for a demonstration of Brillouin lasing. 

When pumped at 1560.5~nm, we observed Brillouin lasing oscillation in the backward direction in the nearest mode, TE6, downshifted by 8.1~GHz from the pump frequency (Fig.\ref{fig3}g). We further characterized the slope efficiency of our Brillouin laser by varying the pumping power. The result is shown in Fig.\ref{fig3}h, with an estimated threshold power of 47~mW and a slope efficiency of 2\%. When compared to previously demonstrated Brillouin lasers in chalcogenide, silicon and silicon nitride cavities \cite{Morrison_Optica_2017,Otterstrom2018ALaser,Ji_Science_2024}, our laser exhibits a much reduced footprint. In terms of the laser efficiency, our current results are limited by the cavity loss and the coupling efficiency to the bus-waveguide of the Stokes wave residing in the TE6 mode. Through improved design, we estimate that a threshold power on the order of a few milliwatts, allowing for on-chip laser pumping, and slope efficiency of 30 percent can be achieved. This will bring our laser to comparable power performance as Brilloun lasers in competing technologies with two orders of magnitude hardware compression, whilst the intermodal nature prevents cascading Brillouin lasing that normally limits the output power and noise performance in intramodal counterparts \cite{Li_ncomms_2013,Gundavarapu_NatPhot_2018,Behunin_PRA_2018}.

We investigated the phase noise performance of our intermodal Brillouin laser. In contrast to previous demonstration in silicon nitride, where optical cavity loss dissipation is much lower than acoustic dissipation that leads to the operation regime of Stokes linewidth narrowing \cite{Blumenthal_Book_2022}, our laser operates in the phonon lasing regime \cite{Otterstrom2018ALaser}, where the long lifetime of the SAW exceeds the photon lifetime in the cavity by an order of magnitude. To date this regime is inaccessible in any silicon nitride based Brillouin laser. The measured single-sideband phase noise ($\mathcal{L}(f)$) of the Brillouin laser along with the pump laser is depicted in Fig.\ref{fig3}i. Through fitting \cite{Li_ncomms_2013,IEEE_std_Phasenoise_2021} we estimate the intrinsic linewidths of the pump and Brillouin lasers to be 628 Hz. Due to phonon lasing operation, these lasers obtained the same fundamental linewidth \cite{Otterstrom2018ALaser}. We observed significant linewidth narrowing in the RF beat note between the pump and Brillouin laser due to common noise cancellation of the two light waves mediated by the phonons. In this case a record low intrinsic linewidth of 7.1~Hz was observed, which would only otherwise be possible in large ultra-low loss cascaded Brillouin lasers \cite{Li_ncomms_2013,Gundavarapu_NatPhot_2018}. This opens the potential of our device to directly function as a compact low noise RF oscillator. 

\begin{figure}
\centering
\includegraphics[width=\linewidth]{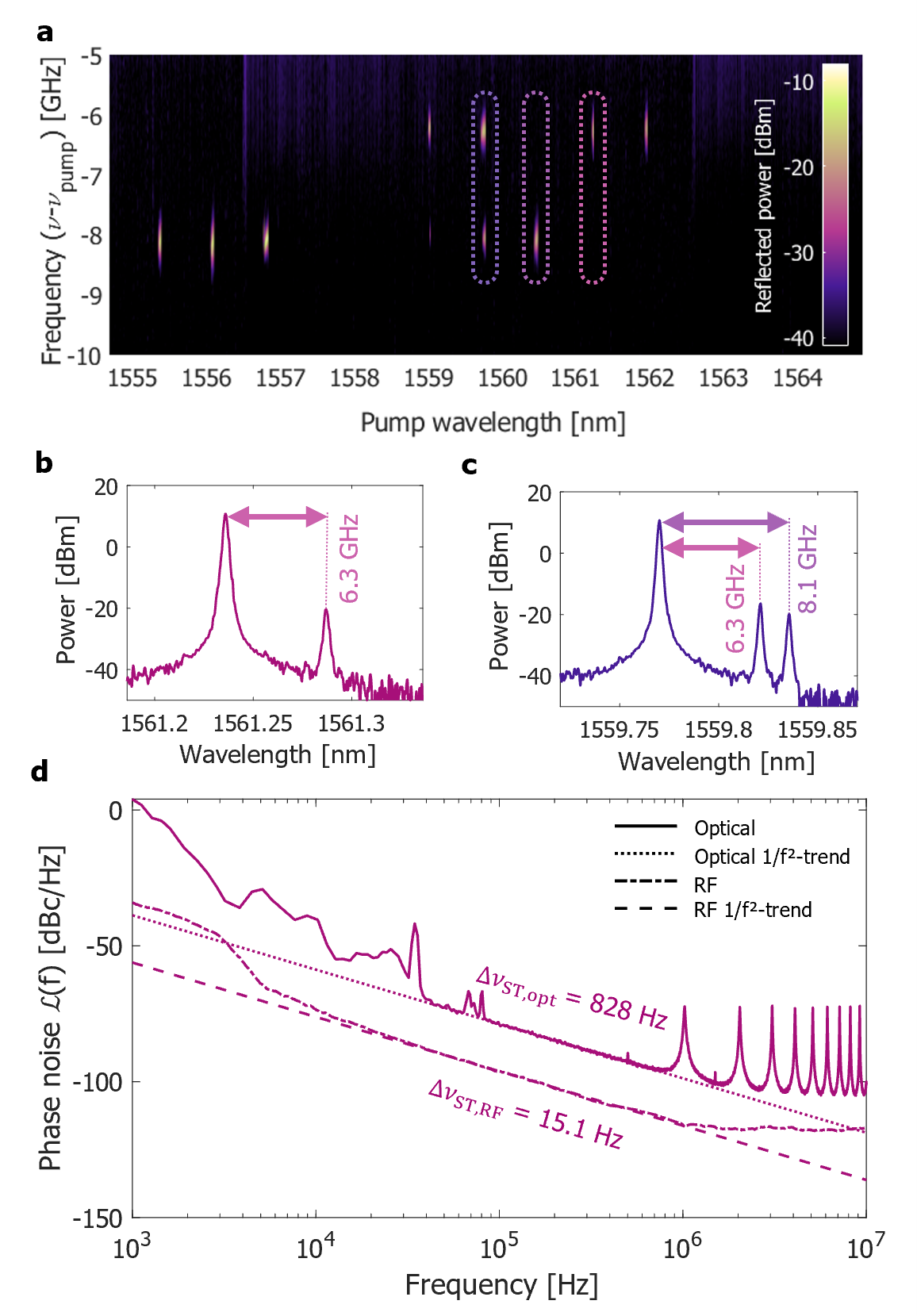}
\caption{\textbf{Multiwavelength operation of the Brillouin laser.} \textbf{a} Reflected optical spectrum as a function of increasing wavelength of the pump. Encircled are operation points with spectra shown in separate figures (from left to right): Fig.\ref{fig4}c, Fig.\ref{fig3}g and Fig.\ref{fig4}b. \textbf{b} Reflected optical spectrum recorded of the bulk acoustic wave Brillouin laser. \textbf{c} Reflected optical spectrum recorded of the Brillouin laser in multi-wavelength operation. \textbf{d} Phase noise $\mathcal{L}(f)$ measured for the bulk acoustic wave mediated Brillouin laser and the beatnote between reflected pump and Brillouin laser.}\label{fig4}
\end{figure}

\begin{figure*}
\centering
\includegraphics[width=\textwidth]{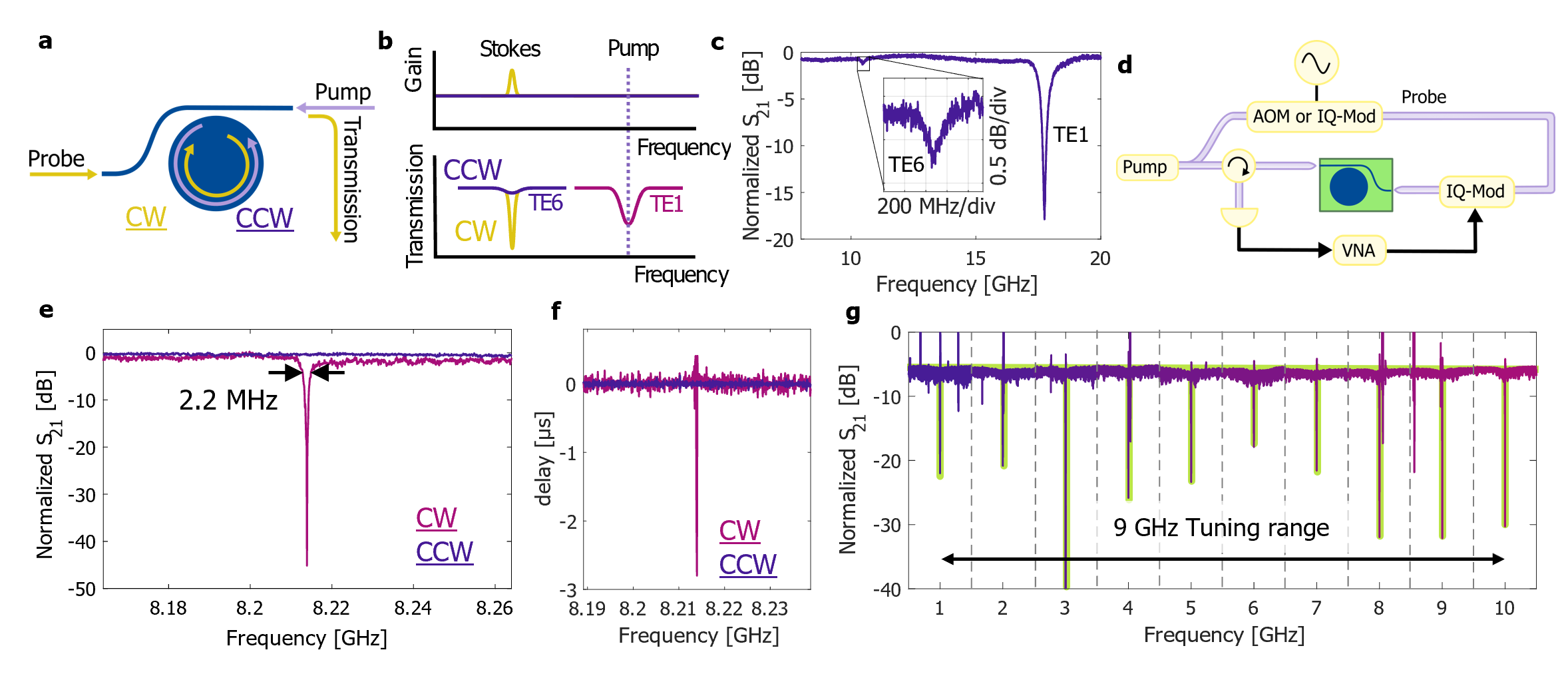}
\caption{\textbf{Non-reciprocity, Brillouin induced opacity and tunable microwave photonic notch filter.} \textbf{a} Principle of the pump probe test for non-reciprocity where the pump affects the transmission of the probe only when counter propagating (CW). \textbf{b} Schematic of the transmission and gain spectrum for the Brillouin induced opacity. \textbf{c} Transmission spectrum ($S_{21}$) of the TE1 and TE6 mode. \textbf{d} Schematic of the Brillouin induced opacity demonstration using an acousto-optic modulator (AOM) and for the demonstration of a tunable filter and IQ-modulator. \textbf{e} Transmission spectra of the probe in the counter (CW) and co-propagating (CCW) direction with respect to the pump, centered around TE6. \textbf{f} Group delay calculated from the phase response corresponding to the transmission in e. \textbf{g} RF transmission spectrum of the Brillouin induced opacity based filter, with 1 GHz bands for separate measurements tuning the filter by 1 GHz steps over 9 GHz. Green highlight illustrates the locations of the baseline and  filter notches.}\label{fig5}
\end{figure*} 

\subsection*{Multi-wavelength Brillouin laser}\label{secMMSBL}
We show for the first time multi-wavelength Brillouin lasing in the resolved sideband regime, enabled by the unique dual mode characteristic of our waveguide. Through a simple pump frequency tuning mechanism in the same device, we have a single or dual wavelength lasing states as shown in Fig.\ref{fig4}a. Compared to the SAW based Brillouin laser we find similar noise performance of the Brillouin laser operating at the 6.3~GHz bulk acoustic mode, see Fig.\ref{fig4}b,d, where the peak displacement in this mode is contained within the tellurite layer. We further study the behaviour of the laser in multi-wavelength operation as shown in Fig.\ref{fig4}c, and find the Brillouin shift to match the acoustic SAW and bulk mode. Dual wavelength Brillouin lasing has novel applications for enabling ultra sensitive cavity thermometry \cite{Loh_optica_2019} and RF signal generation with long term stability \cite{Mercade_PRL_2021}.

\subsection*{Brillouin-induced opacity}
The directionality of Brillouin scattering enables novel non-reciprocal integrated photonic devices. Particularly, in a cavity it is possible to enable Brillouin induced transparency and opacity \cite{Kim_nphys_2015,Dong_ncomms_2015}, but this has so far not yet been demonstrated in planar integrated photonic platforms. We observe Brillouin induced opacity in our microdisk cavity by pumping the Brillouin laser just below threshold, creating constructive amplitude coupling of the probe to the resonator. This allows the strongly undercoupled resonance with a shallow extinction to get to a near critically coupled state at the Brillouin gain peak, creating very high light extinction. This effect is know as Brillouin induced opacity and is a non-reciprocal effect, where a probe light counter propagating with the pump will experience high extinction, while the co-propagating light at the same frequency will be transmitted (Fig.\ref{fig5}a-c).

Moreover, the bandwidth of the Brillouin induced opacity is governed by the interference effect of the cavity resonance with the Brillouin gain resonance. This will give rise to a Fano resonance effect that experiences significant linewidth narrowing. We observe (see methods, (Fig.\ref{fig5}d)) an ultra-narrow linewidth of 2.2~MHz from our induced opacity and a strong extinction of more than 45 dB, as depicted in Fig.\ref{fig5}e. Such linewidth narrowing represents an order of magnitude reduction when compared to the measured Brillouin linewidth of our waveguides. When measuring the phase response of the opacity resonance we observe advancement (fast light) of 2.8 microseconds (Fig.\ref{fig5}f), which is a three-orders of magnitude enhancement when compared to Brillouin slow or fast light effects in our single pass waveguides.

The Brillouin induced opacity generated in the disk makes for an excellent MWP notch filter which is tunable over a wide 9~GHz range by simply changing the carrier frequency of the light, as demonstrated in Fig.\ref{fig5}g. To the best of our knowledge, our results represent the narrowest linewidth ever demonstrated in integrated microwave photonic filters. When measured at the narrowest point, namely, 3~dB above the minimum transmission point, our filter exhibits an exceptionally narrow bandwidth of \textless{}~25~kHz, which is an order of magnitude finer that the value reported using SBS in spun fiber \cite{Choksi_nphot_2022}. Along with the wide tuning range and the simple single sideband modulation scheme, our opacity-based filter demonstration shows clear advantage over previous demonstrations \cite{Marpaung_Optica_2015,Gertler_ncomms_2022, Matthew_ncomms_2023}.

\section*{Discussion}
In this work we demonstrate strong Brillouin scattering in a silicon nitride based platform by introducing a layer of tellurite produced in a single deposition step. The Brillouin gain coefficient increased to 81~m$^{-1}$W$^{-1}$ representing a significant two orders of magnitude improvement of SBS strength in silicon nitride. This enables a wide range of technologies and applications previously inaccessible in scalable standard silicon nitride platforms.

Utilizing a single pass waveguide, we have shown for the first time net gain Brillouin amplifiers in silicon nitride, along with slow light tunable delay and a narrowband MWP notch filter. The strong SAW Brillouin interaction also allows for the design of an intermodal Brillouin laser in compact planar microdisks, where the long phonon lifetime drives the device to function directly as an RF oscillator with a narrow linewidth of 7~Hz. In addition the first-ever demonstration of multi-wavelength Brillouin lasing, may also lead to precise cavity thermometry \cite{Loh_optica_2019} and long term stabilized signal generation \cite{Mercade_PRL_2021}. Finally utilizing Brillouin induced opacity in the microdisk we created the narrowest tunable MWP notch filter with spectral resolution of 2.2~MHz.

Bringing this technology to silicon nitride allows for harnassing powerful SBS signal processing in a scalable and ultra-low loss platform and intersecting SBS with complementary functionalities such as rare-earth-doped segments for amplifiers \cite{Frankis_PhotRes_2020} and on-chip pump lasers \cite{Kiani_OE_2023}.

\section*{Methods}
\subsection*{Waveguide fabrication}
The silicon nitride waveguide were fabricated on a 100~mm diameter silicon wafer using a standard foundry process \cite{Roeloffzen_JSTQE_2018}. This involves growing a 6-micron silicon oxide layer through wet thermal oxidation of the silicon substrate at 1000°C. A 200~nm-thick silicon nitride layer was then grown using low-pressure chemical vapor deposition (LPCVD). The wafer was annealed in nitrogen at \textgreater{} 1100 °C for several hours to drive out hydrogen from the Si3N4 layer. The waveguides were patterned into the silicon nitride layer by stepper lithography and reactive ion etching. The wafer was then diced into individual chips. Next the 292~nm-thick tellurite layer, was grown using radio frequency (RF) reactive sputtering \cite{Frankis_OptExp_2019}. For the silica cladded samples as used in the notch filter, an additional 970~nm-thick silicon oxide layer was deposited using plasma-assisted reactive magnetron sputtering (PARMS) \cite{TorabAhmadi_FCSE_2024,TorabAhmadi_JVSTA_2024}. The waveguide based devices used a 2-µm (1.6-µm) wide silicon nitride core for the unclad (silica cladded) geometry.

\subsection*{Brillouin gain measurement}
The gain coefficient measurements were performed using an updated version of the double intensity modulation setup previously used in our work on multilayer silicon nitride waveguides \cite{Ye_APL_2023,Botter_SciAdv_2022}, which was based on earlier work in Brillouin spectroscopy \cite{Grubbs_RSI_1994} and gain measurements in thick silicon nitride waveguides \cite{Gyger_PRL_2020}. In this setup both the pump and probe laser are modulated at slightly different frequencies. This way any interaction between them when traveling through the sample (i.e., SBS) will result in a signal at their difference frequency such that the interaction strength can be recorded using a lock-in amplifier.

\subsection*{Amplifier, delay and filter measurement}
In the amplifier, delay and filter experiments, a VNA based measurement setup was utilized as schematically indicated in Fig.\ref{fig2}b. Here a pump laser is coupled to the chip using a lensed fibres, whilst in the counter-propagating direction a probe laser is injected. The probe laser is set at a lower frequency and an upper sideband modulation is introduced using an in-phase quadrature (IQ) modulator controlled by a VNA. The modulation creates an RF to optical map which is retrieved back to the RF in the transmission in at a photodiode, retrieving with the VNA both optical phase and amplitude information of the sideband. This sideband scans over the Brillouin gain at various pump powers which is used to determine the off and on resonance transmission to determine the on-off gain. The delay was calculated at this peak gain from the phase response as $\tau=-\frac{d\phi}{d\omega}$.

For the microwave photonic notch filter, the IQ modulator is biased differently as mentioned in the main-text where the sideband phase and amplitudes are tailored for the Brillouin gain to create destructive interference of the RF signal at the photodetector. In this demonstration a different geometry with a silica top-cladding on the tellurite layer was used instead of the uncladded geometry used elsewhere in the article. 

\subsection*{Brillouin laser phase noise measurements}
In the RF phase noise measurement of the beatnote of the reflected pump reflections of the waveguide facet and the Brillouin laser output, the light returned from the chip is directly sent to the photodiode. The output of the photodiode is either directly analysed using an electrical signal analyzer with a phase noise measurement extension, as is done for Fig.\ref{fig4}d, or it is recorded using a high-speed oscilloscope (13 GHz) from which the phase deviation over time is used to calculate the phase noise power spectral density (PSD), as is done for Fig.\ref{fig3}d.

For the optical phase noise measurements we use a delayed self-heterodyne detection system. First the pump is filtered out from the reflected signal using a tunable bandpass filter, after which it is sent into an imbalanced Mach-Zehnder interferometer (iMZI), with one arm containing 200~m additional fibre and the other arm applies an 80~MHz frequency shift using an AOM. The iMZI output is recorded using a balanced photodiode and a high-speed oscilloscope. From the recorded frequency deviation we calculate the single sideband frequency noise PSD, after which it is converted to the optical frequency noise using the gain function of the transfer of the frequency noise \cite{Yuan_OE_2022}. Finally, the single sideband frequency noise PSD $S_f^I(f)$ is converted to the phase noise $\mathcal{L}(f)$ \cite{IEEE_std_Phasenoise_2021}.

\subsection*{Brillouin-induced opacity}
For the Brillouin induced opacity demonstration, a VNA based measurement setup was utilized as schematically indicated in Fig.\ref{fig5}d. A pump laser is coupled to the TE1 mode of the resonator and a probe is coupled in the counter-propagating (co-propagating) direction of the disk to measure the Brillouin induced opacity (unaffected TE6  resonance). The probe laser is a frequency shifted copy of the pump generated using an AOM (80~MHz), to increase the relative stability of the pump and probe laser needed to resolve the incredibly narrow transmission features. The probe is then modulated with the VNA controlled RF signal using an IQ-modulator in a singlesideband (lower sideband) configuration to characterize the frequency band around the pump. This the transmission of the probe is measured using a fast photodiode (\textgreater{} 23 GHz) and recorded using the VNA to create the $S_{21}$ transmission spectrum.

The tunable microwave photonic notch filter is created by the Brillouin induced opacity, where the offset of the probe carrier from the pump is used to control the filtering frequency. In this case we use the same setup except the probe laser is now shifted using an IQ-modulator biased to be in a suppressed carrier single sideband state instead of an AOM. The offset is controlled by an RF signal synthesizer. The probe laser is now down shifted by 9.123 to 18.123 GHz from the pump using an IQ-modulator, and the RF signal is encoded on the light as the upper frequency sideband, mapping the opacity to the RF when recording the probe on the photodetector.

\section*{Author contributions}
% V1: 2 sentences of McMaster specific contribution
%Y.K. and D.M. developed the concept and proposed the physical system. R.M., Y.K., R.A.B. and K.Y. developed and performed numerical simulations and optimization. Y.K., A.M., R.A.B. and R.M., performed the experiments with input from K.Y., D.M., R.B.G.B., P.J.M.S., B.L.S.F. and B.H.. B.H., B.L.S.F. and H.M. contributed to the waveguide and resonator design with input from J.D.B.B.. B.H., B.L.S.F., P.T.A., N.M.T. and E.J. performed platform development, device fabrication and device validation under supervision of P.R.S, P.M. and J.D.B.B.. D.M. and Y.K. wrote manuscript with contributions of all authors. D.M. supervised the project.

% V2: 1 sentence of McMaster specific contribution
Y.K. and D.M. developed the concept and proposed the physical system. R.M., Y.K., R.A.B. and K.Y. developed and performed numerical simulations and optimization. Y.K., A.M., R.A.B. and R.M., performed the experiments with input from K.Y., D.M., R.B.G.B., P.J.M.S., B.L.S.F. and B.H.. B.H., B.L.S.F., H.M., P.T.A., N.M.T., and E.J. designed, fabricated and validated the devices under supervision of P.R.S, P.M. and J.D.B.B.. D.M. and Y.K. wrote manuscript with contributions of all authors. D.M. supervised the project.

% Catagories: 
% Concept and physical system
% Numerical simulations and optimization
% Gain characterization and application demonstrations
% Design, Fabrication and film characterization
% Writing
% overall supervision

%D.M. and R.A.B. developed the concept and proposed the physical system. R.A.B., R.M., K.Y. and Y.K. developed and performed numerical simulations. R.M. performed the numerical optimization with input from K.Y. and Y.K.. Y.K. and R.A.B. performed the gain characterization experiments, with input from R.M., B.L.S.F., B.H. and K.Y.. A.M. performed the microwave photonic notch filter experiment with input from D.M. and Y.K.. Y.K. performed the Brillouin amplifier and delay measurement. Y.K. proposed and performed the Brillouin laser experiments with input from D.M., R.M., P.J.M.S. and non-reciprocity with input from D.M. and R.B.G.B..B.H., B.L.S.F. and H.M. contributed to the waveguide and resonator design with input from J.D.B.B.. B.H, B.L.S.F. and P.T.A. carried out the TeO$_2$ and SiO$_2$ thin film sputtering fabrication steps. B.H., N.M.T. and E.J. characterized the passive optical properties of the thin films and tellurite-silicon nitride waveguides and resonators. P.R.S. and P.M. contributed to funding aquisition and supervision and J.D.B.B. supervised the tellurite-silicon nitride platform development and coordinated the collaboration with D.M.. D.M. and Y.K., wrote the manuscript with contributions from R.A.B. and R.M.. D.M. supervised the project. 

% NEEDS TO BE MORE COMPACT! YOU SHOULD HAVE SEEN THE POOR CONDITION OF THIS SEGMENT!

\section*{Acknowledgement}
The authors thank Centre of Emerging Device Technologies (CEDT) for support with tellurite depositions, Intlvac Thin Film for silicon dioxide deposition and the authors would like to thank Lisa Winkler and Albert van Rees for helpful discussions on laser noise characterization. 

% Growth fund?
\section*{Funding Information}
The authors acknowledge funding from the European Research Council Consolidator Grant (101043229 TRIFFIC) and Proof of Concept grant (101157112 Veritas), Nederlandse Organisatie voor Wetenschappelijk Onderzoek (NWO) Vidi (15702) and Start Up (740.018.021), the Photon Delta National Growth Fund programme, the Natural Sciences and Engineering Research Council of Canada (NSERC) (I2IPJ 555793-20 and RGPIN-2017-06423), the Canadian Foundation for Innovation (CFI) (35548), Mitacs (IT23542 FR60781) and the McMaster University Faculty of Engineering (Doctoral Scholarship for Multi-disciplinary Research).

\bibliographystyle{IEEEtran}
\bibliography{library}

\end{document}